\documentclass[conference]{IEEEtran}

\usepackage[pdftex]{graphicx}
\graphicspath{{../pdf/}{../jpeg/}}
\DeclareGraphicsExtensions{.pdf,.jpeg,.png}
\usepackage{tabularx}
\usepackage{booktabs}
\usepackage[labelfont=bf,format=plain,justification=raggedright,singlelinecheck=false]{caption}
\usepackage[cmex10]{amsmath}
\usepackage{mathabx}
\usepackage{graphicx}
\usepackage{subcaption}
\usepackage{floatrow}
\usepackage{algorithmic}
\usepackage{array}
\usepackage{mdwmath}
\usepackage{caption}
\usepackage{mdwtab}
\usepackage{eqparbox}
\usepackage{url}
\begin{document}

\title{\LARGE
Mitigating Adversarial Effects of False Data \\ Injection Attacks in Power Grid
}

 \author{
     \authorblockN{
         Farhin Farhad Riya\authorrefmark{1}, 
         Shahinul Hoque\authorrefmark{1}, 
         Jiangnan Li \authorrefmark{1}
         Yingyuan Yang\authorrefmark{2}, 
         Hairong Qi\authorrefmark{1} 
         Jinyuan Sun\authorrefmark{1}, 
     }
     
     \authorblockA{
        \authorrefmark{1}Department of Electrical Engineering and Computer Science, University of Tennessee
     }
     \authorblockA{
        \authorrefmark{2}Department of Computer Science, The University of Illinois at Springfield
     }
     \authorblockA{ 
        friya@vols.utk.edu, 
        shoque@vols.utk.edu,
        jli103@vols.utk.edu
        yyang260@uis.edu, 
         hqi@utk.edu
        jysun@utk.edu
    }
 
 }

\maketitle

% https://ieeexplore.ieee.org/stamp/stamp.jsp?tp=&arnumber=9384314

\begin{abstract}
Deep Neural Networks have demonstrated remarkable accuracy in various tasks in recent years. Their effectiveness has also been recognized in power grids for detecting False Data Injection Attacks (FDIA) during critical operations such as state estimation. However, due to the vulnerabilities of DNNs and the unique infrastructure of cyber-physical systems (CPS), attackers can exploit these weaknesses to evade detection. Additionally, the distinctive nature of CPS poses challenges for conventional defense mechanisms against False Data Injection Attacks. In this paper, we propose a DNN framework that incorporates an additional layer, utilizing randomly padded inputs to mitigate the adversarial effects. Our method offers a significant advantage when deployed in a DNN model, it has minimal impact on the model's performance, even with larger padding sizes. To demonstrate the effectiveness of our framework, we conduct simulations using the IEEE 14-bus, 30-bus, 118-bus, and 300-bus systems. Furthermore, we select attack techniques that generate subtle adversarial examples capable of effortlessly bypassing the detection mechanism, in order to validate our framework.

\end{abstract}

\begin{keywords}
Deep Neural Networks, False data injection attack, State Estimation, power grid, cyber-physical-system.
\end{keywords}

% ############################### Introduction ###############################
% ###############################              ###############################

\section{INTRODUCTION}
The power grid is a complex system that interconnects multiple electricity sources with consumers through extensive transmission and distribution networks. To guarantee the security of the electric power infrastructure, control, and management systems are vital for the power grid. According to statistics from the North American Electrical Reliability Council, 11 blackouts have occurred due to abnormalities in the cyber system of supervisory control and data acquisition (SCADA)\cite{sheng2011cyber}.

State estimation is a method used to determine unknown state variables in a power system based on meter measurements conducted by the control center. By analyzing the meter measurement data and power system models, state estimation is employed in system monitoring to accurately determine the state of the power grid. Typically, the results of state estimation are utilized in contingency analysis, which then assists in controlling the components of the power grid. As shown in Figure 1, measurement data from sensors or meters, such as bus voltage, bus power flow, branch power flow, and load profiles, are usually transmitted to a control center (SCADA). The control center then analyzes the received measurement data, estimates the states of the power system, identifies potential contingencies, and transmits the appropriate control signals to the Remote Terminal Units (RTUs) to ensure the reliable operation of the power system.
Cyber-physical systems are typically complex, incorporating various geographically dispersed data sources. Meters or sensors are deployed in the field to collect data from the physical environment, as depicted in Figure 1. Given the distributed nature of the system, ensuring the security of all meters is challenging. Furthermore, the data gathered by the sensors often deviates from the theoretically calculated ideal data due to environmental perturbations, resulting in measurement errors. Exploiting this limitation, attackers can introduce carefully crafted perturbations to the compromised meters' measurements, allowing them to bypass anomaly detection systems.
Numerous studies have examined different scenarios of False Data Injection Attacks (FDIA), and some of these attacks \cite{falsedataattack} \cite{yu2015blind}\cite{liu2016masking} enable the generation of subtle malicious false data that can successfully evade detection systems and cause the estimator to produce incorrect outputs. Critical operations such as Optimal Power Flow analysis, Energy Distribution, and Real-time pricing heavily rely on accurate state estimation, making them vulnerable to various successful false data injection attacks.
Likewise, extensive research has been conducted to develop detection strategies for different attack scenarios. Among the proposed defense techniques, Machine Learning models, particularly Deep Neural Networks (DNNs), have shown superior performance in anomaly detection. However, the vulnerabilities of DNN models to well-crafted perturbations pose critical issues, especially in Cyber-Physical systems like the power grid, which have unique constraints compared to other domains such as computer vision. In state estimation, the sensors used to collect data from the physical environment introduce measurement errors, and well-crafted perturbations mimicking these errors can degrade the performance of well-trained DNN models. In light of these challenges, our research aims to provide a framework that enhances the robustness of DNN models and makes attacks computationally expensive for adversaries.
The key contributions in this paper can be summarized as follows:
\begin{itemize} 
  \item We propose a general framework that can be easily adapted by the prevailing Machine Learning techniques that are utilized in FDIA detection.
  \item The vast distributive nature of CPS requires infeasible labor to guarantee the security of the sensors. We highlight that our framework does not require any specific hardware configuration or re-deployment of the sensors.
  \item We propose a defense method that mitigates the adversarial effect by reconstructing the input samples with multiple combinations of random padding. The multi-combination padding expands the number of data samples which hinders the accuracy digression while the size of the padding increases which is crucial for test cases having less number of meters comparing
  \item Our proposed framework has negligible accuracy drop compared to the vanilla models with negligible computational cost changes. 
  \item We validate the framework through simulation using the IEEE test system, including IEEE 14-bus case, 30-bus case, 118-bus, and 300-bus case. For every case, the favorable outcomes justify the proposed mechanism.
\end{itemize} 
\begin{figure}[htbp]
    \centering
    \includegraphics[width=\linewidth]{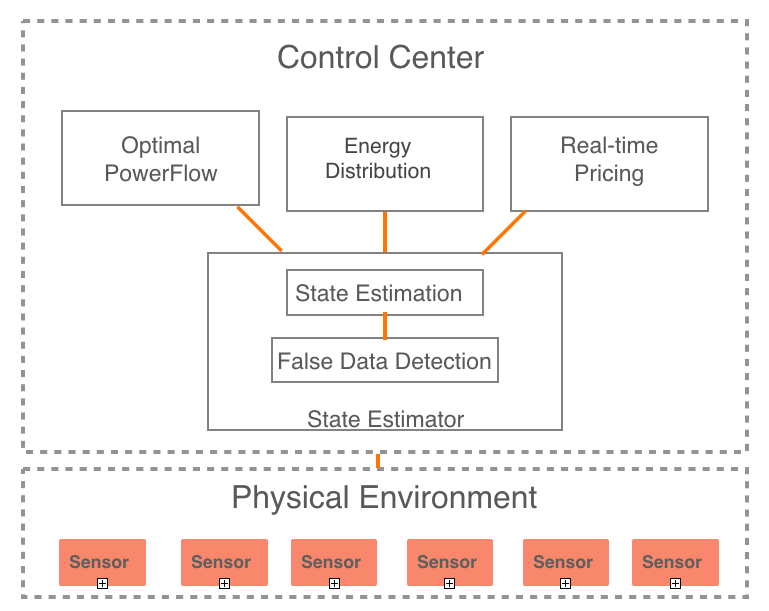}
    \caption{Structure of communication and control system of power grids}
    \label{fig:img1}
\end{figure}

% {\captionsetup{parskip=0pt}
% }
The rest of the paper is structured as follows. Section II provides an overview of the relevant research conducted in the areas of both attack and defense. In section III, we present essential background information and technical details. A subsection of section IV explores different approaches for identifying a suitable attack strategy that effectively justifies the defense mechanism using subtle perturbations. Furthermore, section IV provides a comprehensive explanation of the proposed framework. In section V, we present information about the data set used, the model architecture, and the evaluation results. Section VI delves into future work considerations, and we conclude the paper in Section VII.

% ###############################            ###############################
% ############################### Background ###############################
% ###############################            ###############################
\section{Related Work}
Since 2009, when Y. Liu et al. \cite{liu2009false} revealed the vulnerability of SCADA systems to malicious fake data injection (FDI) attacks, numerous research efforts have been undertaken to understand the potential of FDI attacks due to the critical role that state estimation plays in the power system. Many attack studies have focused on stealthy techniques that surpass the system's residual-based bad data detectors by exploiting their knowledge of the topological Jacobian matrix. Exploiting the sparse nature of the Jacobian matrix, several attack strategies have been proposed, including those in \cite{kosut2011malicious}, \cite{kim2013topology}, and \cite{ozay2013sparse}, which have shown promising evaluation results. Furthermore, even without full knowledge of the Jacobian matrix, FDI attacks can still be constructed, as demonstrated in \cite{rahman2012false} and \cite{srivastava2013modeling}. The results of some attack studies, such as \cite{yan2016power}, have revealed the potential for stability damage in the power system. C. Jiongcong et al. \cite{jiongcong2016impact} conducted a static security evaluation, examining the effects of FDIA, and concluded that it can lead operators to make incorrect decisions. The effects of both random and structured bad data on state estimation were analyzed in \cite{tajer2014energy}. Data leakage and false injection attacks remain primary concerns for smart grid security since the stability of the grid heavily relies on the integrity of the data \cite{liu2009false}.

To defend against FDI attacks, various defense mechanisms have been studied, focusing on protecting carefully selected measurements \cite{huang2011defending} \cite{bi2011defending}. These mechanisms offer optimal and simple algorithms to preserve data integrity and support their claims with physical experiments. Huang et al. \cite{bi2011defending} proposed an adaptive cumulative sum approach to enable quick detection of false data. Studies such as \cite{kim2011strategic} and \cite{chen2006placement} utilized synchronized PMU measurements with GPS signals.

With the increasing popularity of machine learning (ML) models in various fields, different studies have been conducted in the power grid domain, leveraging the advantages of ML models for their performance and ease of implementation. Unlike the aforementioned defenses, ML models do not require additional hardware devices or sensor redeployment. Furthermore, ML models eliminate the need to solve complex time-domain equations related to power grid analysis, making the strategies computationally efficient. Machine learning has become one of the primary methods used for detecting FDIA. Ozay et al. \cite{ozay2016machine} generated Gaussian dispersed attacks to classify attacks and employed both supervised and semi-supervised ML techniques. Similarly, Esmalifalak et al. \cite{esmalifalak2014detecting} developed an unsupervised learning case-based statistical anomaly detection approach and a distributed SVM-based model for labeled data. \cite{deng2018false} utilized Recurrent Neural Networks (RNN) for FDIA detection. In order to capture the dynamic behavior of the power system, \cite{hochreiter1997long} employed a recurrent neural network with LSTM cells, while \cite{krizhevsky2012imagenet} used a convolutional neural network to balance two input sources. Several studies have employed Deep Neural Networks (DNN) to defend against FDIA \cite{mohammad2018detecting} \cite{JQ2018Online} \cite{AS2019Survey}.

% \begin{figure*}
% \centering 
%     \includegraphics[width=0.8\columnwidth]{figure/framework.png}
%     \caption{Overview of the proposed framework}
%     \label{fig:framework}
% \end{figure*}

Although DNN models have demonstrated excellent results among ML models, they are also susceptible to various attacks. Szegedy et al. \cite{szegedy2013intriguing} uncovered the existence of adversarial scenarios for DNNs in 2013. DNNs can be manipulated to generate false results by adding purposefully created perturbations to the inputs. Several adversarial attack techniques have been proposed, including the Fast Gradient Sign Method (FGSM) presented by Goodfellow et al., which generates perturbations using signed gradient values \cite{goodfellow2014explaining}. Rozsa et al. developed the Fast Gradient Method, which directly utilizes the gradient values \cite{rozsa2016adversarial}. Other adversarial attack methods include iterative attacks \cite{kurakin2016adversarial} and DeepFool \cite{moosavi2016deepfool}. Various defense techniques have been proposed to mitigate these attacks, such as adversarial training \cite{kurakin2016adversarial}, model distillation \cite{papernot2016distillation}, adversarial detection \cite{xu2017feature}, and input reconstruction \cite{gu2014towards}.

% \begin{figure*}
%     \centering
%     \includegraphics[width=0.8\textwidth]{figure/MLCanary_design_2.png}
%     \caption{Verification of model using MLCanary.}
%     \label{fig:img2}
% \end{figure*}

\section{Background}
\subsection{State Estimation in Power System}
In this section, we will provide a brief overview of DC state estimation, including the detection of bad data and the false data injection attacks discussed in [1]. The notation used throughout this section is summarized in Table 1.

A common formulation of the state estimation problem when using a DC power flow model. The relationship between the sensor measurements and the true states of the system can be expressed as:

\begin{equation}
    z = Hx + e \label{eq:1}
\end{equation}

Here, $x = (x1, x2, ..., xn)^T$ represents the true states of the system that we aim to estimate, and $z = (z1, z2, ..., zm)^T$ represents the sensor measurements. The matrix H is an m × n Jacobian matrix, where Hx is a vector of m linear functions that link the measurements to the states. The vector $e = (e1, e2, ..., em)^T$ represents random errors in the measurements.

The observability of a power network refers to its ability to estimate the states based on the available measurements. Various sensor placement algorithms exist to determine the set of measurements required for observability. Typically, there are more sensors in the power network than necessary for observability, i.e., $m > n$. The minimum set of measurements needed to estimate the n-state variables is known as the set of basic measurements or essential measurements. The remaining measurements are referred to as redundant measurements. These redundant measurements play a crucial role in identifying erroneous sensor measurements.

For DC state estimation, any set of n measurements with linearly independent corresponding rows in H is sufficient to solve for the n state variables and can be considered as the set of basic measurements. In other words, n independent linear equations are adequate to solve for n variables. However, in cases where m is greater than n (which is typically true), state estimation involves solving an over-determined system of linear equations. A commonly used approach is to solve it as a weighted least squares problem, which leads to the following estimator:

\begin{equation}
   \hat{x} = (H^TWH)^{(-1)}H^TWz \label{eq:2}
\end{equation}

Here, W is a diagonal matrix with the elements representing the weights assigned to the measurements. The weights in W are often determined based on the reciprocals of the measurement error variances. It is worth noting that assuming the sensor measurement errors to follow a normal distribution with zero mean, other commonly used estimation criteria such as maximum likelihood and minimum variance also lead to the same estimator given in (2).

\textit{Bad Data Detection:}
% Bobba
The accuracy of sensor measurements used in state estimation can be compromised due to various factors such as device misconfiguration, device failures, malicious actions, or other errors. These inaccuracies can have a detrimental impact on the estimation of state variables. As a result, it holds immense importance for power system operators to actively identify and detect the presence of erroneous measurements in order to ensure the reliability and precision of the state estimation process. 
Numerous methods have been proposed to detect, identify, and rectify inaccurate measurements in power systems. One prevalent approach [18], [17] to identify the presence of erroneous data involves analyzing the L2-norm of the measurement residual. The measurement residual represents the difference between the observed measurements and the estimated measurements. By examining the L2-norm of this residual, potential discrepancies or anomalies in the data can be detected.

\begin{equation}
   ||z-H\hat{x}||\label{eq:3}
\end{equation}

In equation (3), the state estimate is denoted as $\hat{x}$, and the measurement residual is represented as the difference between the vector of observed measurements and the estimated measurements, $z-H\hat{x}$. Intuitively, when the observed measurements (z) contain inaccurate or unreliable data, the L2 norm of the measurement residual tends to be high. Therefore, if the value of the expression in equation (3) exceeds a specific threshold ($\tau$), it is assumed that the data is flawed. 
Under the assumption of mutual independence among all state variables and normal distribution of sensor errors, it can be proven that the squared L2-norm of the measurement residual, $(||z-H\hat{x}||)^2$, follows a chi-squared distribution with degrees of freedom $\nu$ = m - n, where m represents the number of measurements and n represents the number of state variables [18]. By conducting a hypothesis test with a significance level $\alpha$, the threshold value ($\tau$) can be determined.

\subsection{False Data Injection Attack}
The False Data Injection Attack (FDIA) is a method that enables an attacker to generate a perturbed measurement vector, denoted as a = $[a1, a2, ..., am]^T$, which is then added to the legitimate measurement vector z. The resulting measurements become polluted, forming $z_a = z + a$. An academic reference [1] reveals that if the attacker possesses knowledge of the matrix H, they can construct a specific value for a, represented as $H_c$, that has the ability to bypass fault detection in state estimation. Equation (4) illustrates this concept, where $\hat{x}_bad$ and $\hat{x}$ represent the estimated state vectors obtained using $z_a$ and z, respectively.

\begin{equation}
   ||z_a-H\hat{x}_bad|| = || z+a-H(\hat{x}+c) ||\label{eq:4a}   
\end{equation}

The equation (4) can be further simplified as (5) and (6). These equations demonstrate that the norm of the difference between $z_a$ and $H\hat{x}_bad$ is less than or equal to a predefined threshold ($\tau$), thereby evading detection.

\begin{equation}
   || z-H(\hat{x}+(a-h_c)) || \label{eq:4b}
\end{equation}

\begin{equation}   
   || z-H(\hat{x} \leq (\tau) || \label{eq:4c}
\end{equation}

% ∥za − Hˆxbad∥ = ∥z + a − H(ˆx + c)∥ (4a) 
% ∥z − Hˆx + (a − Hc)∥ (4b)
% ∥z − Hˆx∥ ≤ τ (4c)

Furthermore, equation (7) in the same scholarly work [1] provides an efficient means to generate a valid vector a. In this equation, matrix P is defined as $P = H(H^T H)^(-1)H^T$, and matrix B is defined as $B=P-I$, where I represents the identity matrix. By solving the equation

\begin{equation}   
   Ba = 0 \label{eq:7}
\end{equation}

it is revealed that a satisfies the homogeneous equation BX = 0. If the attacker compromises k measurements within z, resulting in k non-zero elements in an equation (7) can be rewritten as 

\begin{equation}   
   B'a' = 0 \label{eq:8}
\end{equation}

where B' is an $m \times k$matrix obtained by sampling corresponding columns from B, and a' is an $m \times k$ matrix obtained by sampling corresponding rows from a based on the k compromised measurements. The scholarly reference [1] demonstrates that as long as k exceeds m - n, where n represents the number of columns in matrix H, there will always exist a non-zero solution for a.

% \subsection{State-of-art Defense Method}

\section{Approach}
\subsection{Overview of Generating Adversarial Examples}

The black-box scenario represents the most practical knowledge constraint for an attacker. In this scenario, the attacker can compromise a subset of sensors within the Cyber-Physical System (CPS) and manipulate their measurement data. However, the attacker lacks knowledge of the measurements from uncompromised sensors and cannot modify them. To evaluate the effectiveness of attacks when different subsets of meters are compromised, we conducted experiments using various bus test cases. Table 1 illustrates the different cases we considered during the generation of attack examples.
Moreover, along with the knowledge constraint in practical power systems, a constant data sampling period is typically employed. For instance, traditional SCADA systems have a sampling period of approximately 2 to 4 seconds, while recently developed Phasor Measurement Units (PMUs) maintain a higher data reporting rate ranging from 10Hz to 60Hz [46]. It is crucial for an adversarial attack to be completed within a single sampling period of the target power system. However, there is a concern regarding the loop in line 5 of Algorithm 1, as it may continue executing beyond the sampling period. To address this issue, in our simulations, we set a threshold for the cycling number or the maximum number of iterations the loop can execute. We ensure that the computational cost of the loop does not exceed the duration of the sampling period. Once this threshold is reached, or in other words, when the loop has executed for a maximum number of iterations within the sampling period, we terminate the algorithm and return the generated perturbation v. This approach allows us to control the computational cost of the adversarial attack and ensure its feasibility within the time constraints imposed by the sampling period. By setting a suitable threshold, we strike a balance between the effectiveness of the attack and the practical limitations of the power system.
\begin{table}
\centering
\caption{Table of Notations}
\label{tab:notations}
\begin{tabularx} {\textwidth} {@{}ll@{}}
\hline

$m$ & The number of measurements \\
$n$ & The number of state variables \\
$H$ & m × n Jacobian matrix representing the topology \\
$x$ & n × 1 vector of state variables\\
$z$ & m × 1 vector of meter measurements\\
$e$ & Measurement error vector \\
$h$ & Function relating state variables and measurements \\
$\hat{x}$ & Estimated state vector \\
$W$ & Covariance matrix of measurement errors \\
% $\tau$ & Threshold for polluted measurements \\
\hline
\end{tabularx}
\end{table}

\begin{figure}[htbp]
    \centering
    \includegraphics[width=\linewidth]{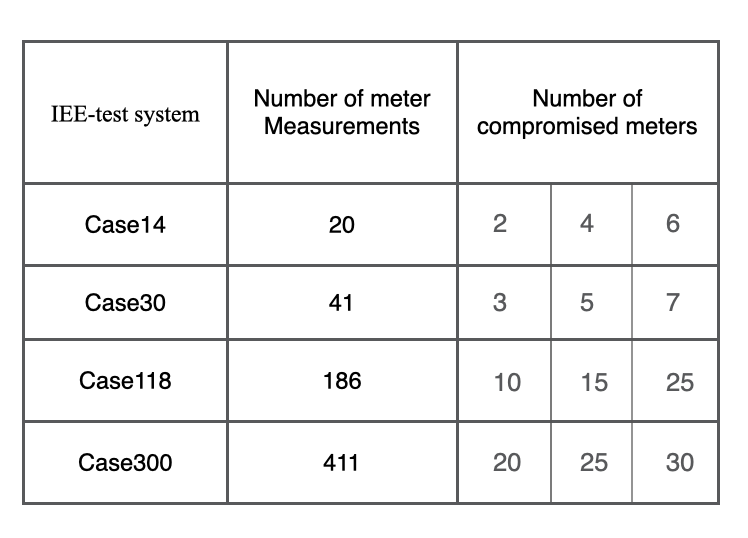}
    \caption{Number of compromised meters in different bus test cases}
    \label{fig:img1}
\end{figure}

\begin{figure}
\centering 
    \includegraphics[width=\linewidth]{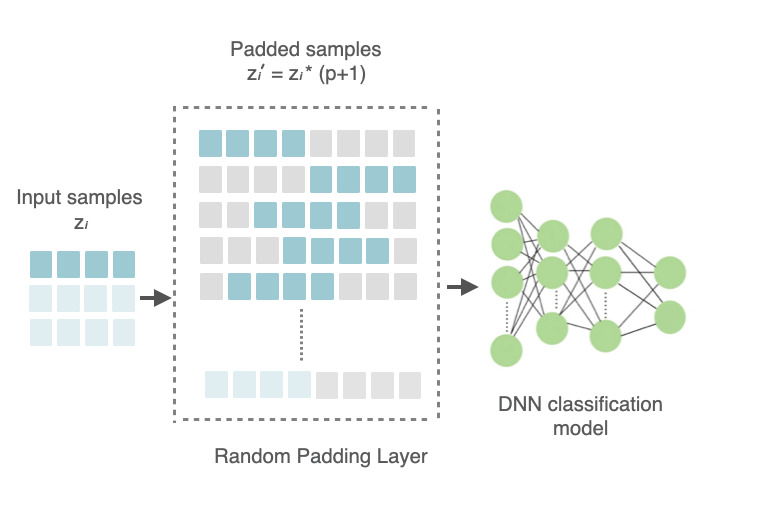}
    \caption{Overview of the proposed framework during training time}
    \label{fig:framework}
\end{figure}

% \begin{figure*}
% \centering 
%     \includegraphics[width=0.8\columnwidth]{figure/new_framework2.png}
%     \caption{Overview of the proposed framework}
%     \label{fig:framework}
% \end{figure*}

\subsection{Overview of Proposed Framework}

The defense against adversarial attacks in the context of DNN-based FDIA (Fault Detection, Isolation, and Accommodation) detection poses a significant challenge. This challenge arises from the fact that adversarial measurements and general false measurements tend to occupy the same manifold, making it difficult to distinguish between them. In order to exploit this vulnerability, attackers employ a gradient-based optimization process to generate perturbations referred to as v specifically targeted at a victim model (za). Interestingly, it has been observed that multi-step attacks often result in perturbations that have limited transferability, indicating that the perturbation v used for FDIA detection is likely to be unique to each victim model. This observation leads to the intuition that the effectiveness of perturbations diminishes when the input to the model is altered. 
To handle this issue, we draw inspiration from stochastic-based defense mechanisms commonly used in computer vision and propose a defense framework called randomization of input padding. This framework aims to mitigate the impact of adversarial attacks in FDIA detection by introducing randomness in the input data. Specifically, our framework involves the addition of a random padding layer before the Deep Neural Network (DNN) during both the training and inference stages. In the training process, sensor measurements (z) are utilized as features for training the detection models. 
To implement the randomization of the input padding framework, an operator is required to select a padding dimension number, denoted as P, which may vary for each bus test case. For example, if a bus test case has x number of features in each meter measurement vector, the selected padding size should exceed 90\% of the feature number $|0.9x|$ to observe a noticeable increase in detection accuracy. Subsequently, P zeros are randomly added to each original input z, resulting in a feature size of (P+m) for each input vector. The padding procedure involves adding all possible combinations of zeros to each input vector. For instance, if the operator chooses a padding size of q, a total of $2^q$ new combinations are generated for each input vector, out of which q+1 combinations are unique as only zeros are padded. This leads to a total of t*(q+1) different possible padding input patterns for the DNN. 
During the training stage, the DNN learns patterns from the plain measurements embedded into the padded inputs. In the inference stage, when a new measurement vector (z) is received, it is randomly padded, resulting in P+1 samples. Among these samples, one is randomly selected for classification. The randomization introduced during the padding process ensures that even if the attacker or operator is aware of the entire framework, they cannot determine the final DNN padded input vectors. Furthermore, as multi-step perturbations exhibit relatively weak transferability, the success rate of adversarial attacks is expected to decrease under the random padding framework. Increasing the value of P intuitively further reduces the success rate of adversarial attacks, thereby enhancing the robustness of the DNN employed for FDIA detection.
It is important to note that our framework, in contrast to the approach described in [43], requires input data pre-processing (padding) during the training stage and cannot be directly applied to a pre-trained model. This is due to the fact that the measurement data from a specific power system must adhere to the manifold defined by the system's physical properties. Reshaping, resizing, or directly sampling the measurement vectors would compromise this adherence. However, the padding scale size does not significantly affect the FDIA detection performance for legitimate data vectors (z) and false measurements (za), as long as an appropriate neural network structure is chosen. Moreover, the computational overhead during the training process introduced by our framework is minimal, and it is compatible with various neural network models.

\begin{figure*}
\centering 
    \includegraphics[width=0.8\columnwidth]{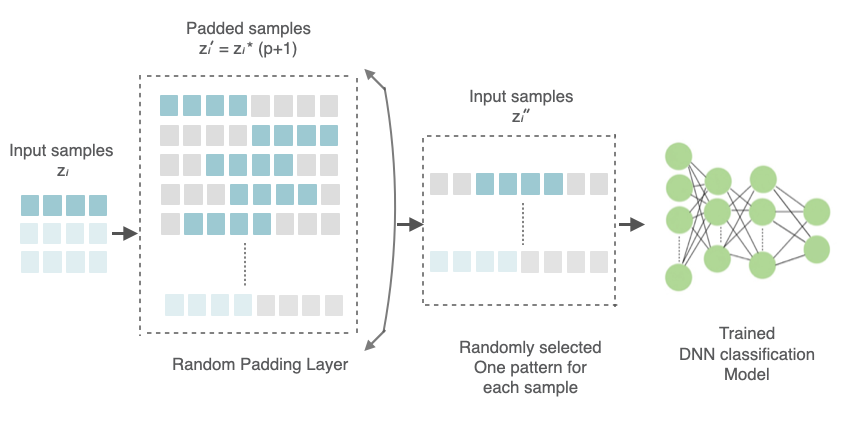}
    \caption{The pipeline of the proposed randomization-based defense mechanism. The input sample z first goes through the random padding layer that pads the samples with all the unique padding combinations z'. From the resulting padded samples for each vector z, one random pattern z" is used for classification.}
    \label{fig:framework}
\end{figure*}

\section{Experiments}
\subsection{Experimental Setup}
\subsubsection{Dataset}
In our simulation, we employed a DC power flow model to study the behavior of power systems. To configure the system for analysis, we utilized MATPOWER, a MATLAB tool specifically designed for addressing power flow problems [35]. Leveraging the capabilities of MATPOWER, we constructed the H matrix, which captures the relationship between the system's state variables and the corresponding meter measurements. By extracting the necessary information from MATPOWER, we were able to derive the meter measurements themselves.
Each measurement vector, denoted as z, contains the power flow measurement data of a particular branch in the power system. The number of measurements within each z vector is denoted as m. It is important to note that the value of m varies for each case, as we considered different IEEE test systems with distinct characteristics and configurations.
In order to obtain a dataset suitable for FDIA (False Data Injection Attack) detection experiments, we generated two classes of data: normal data and false data. This is because FDIA detection is fundamentally a binary classification problem, requiring the ability to differentiate between genuine measurements and maliciously injected data. Our training dataset contains a total of 40,000 samples, covering all the IEEE test cases. Among these samples, approximately 50\% were intentionally perturbed with false data to ensure a diverse and challenging dataset. We labeled the normal data samples as 0, representing the absence of a false data injection, while the false data samples were labeled as 1, denoting the presence of a malicious injection.
To validate the effectiveness of our proposed framework, we conducted multiple experiments across a range of test cases, each with its own unique characteristics. For each test case, we deliberately compromised a different number of meters to assess the impact on the detection results. Through our research, we discovered that compromising approximately 4\% of the total meters had a significant detrimental effect on the accuracy of the detection model, underscoring the vulnerability of power systems to such attacks. To provide a comprehensive evaluation of our framework, we considered various scenarios, Table 1 represents the different scenarios we consider for evaluating our proposed work.

\subsubsection{Defense Model}
In our simulations, we employ a feed-forward neural network F as the target deep neural network (DNN) for training purposes. The structure of the network F is presented in Table II. The number of nodes for the first randomization layer changes according to the selected padding size as the number of features changes for the input samples after adding random padding to the samples.

\begin{table}[h]
  \centering
  \caption{Model Structure of F}
  \begin{tabular}{|c|c|c|c|c|c|c|c|}
    \hline
    \textbf{Layer} & \textbf{1} & \textbf{2} & \textbf{3} & \textbf{4} & \textbf{5} & \textbf{6} & \textbf{7}\\
    \hline
     \textbf{Nodes} & n & 186 & 128 & 64 & 16 & \shortstack{0.25\\Dropout} & 2 softmax \\
    \hline
  \end{tabular}
\end{table}

\subsubsection{Target Model}
The most sophisticated attack strategy involves the attackers meticulously analyzing and considering \textbf{ALL} conceivable patterns of the defense models when crafting adversarial examples. This approach aims to exploit any vulnerability that may exist within the defense mechanism. By taking into account all potential patterns, the attackers seek to ensure the effectiveness of their adversarial attacks.
Nevertheless, such an all-encompassing exploration of defense patterns poses a significant computational challenge. Attempting to target and disrupt a large number of patterns simultaneously demands an exceedingly long period of time to execute, surpassing the computational capacity available to most attackers. Furthermore, there is a risk that this complex and resource-intensive process may not even converge, rendering the entire endeavor futile and ineffective. As a more feasible alternative, an adaptive approach is adopted, whereby the attackers utilize the target models themselves to generate adversarial examples. This approach provides a practical solution to the computational burden while maintaining the capacity to cause considerable harm to the target models. By exploiting the vulnerabilities and peculiarities of the target models, the attackers can create adversarial examples that are highly effective at bypassing the defenses put in place.
 Therefore, we let the attackers utilize the target models for generating adversarial examples by exploring two distinct attack scenarios, enhancing the success rate of their attacks while avoiding the computational in-feasibility of the previous all-encompassing approach.

\begin{figure*}
\centering 
\subfloat{
    \includegraphics[width=.24\textwidth]{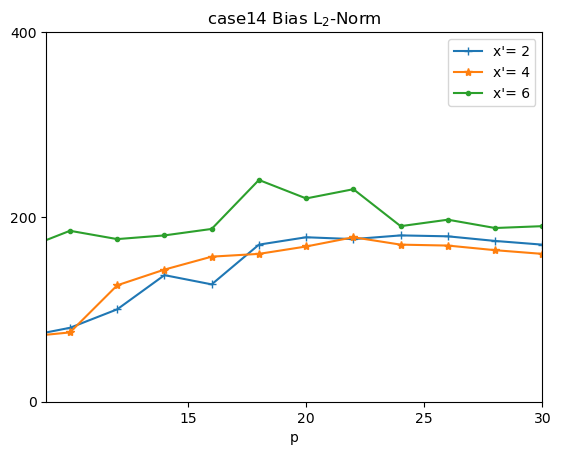}
    \label{fig:case14_bias}
    }
\subfloat{
    \includegraphics[width=.24\textwidth]{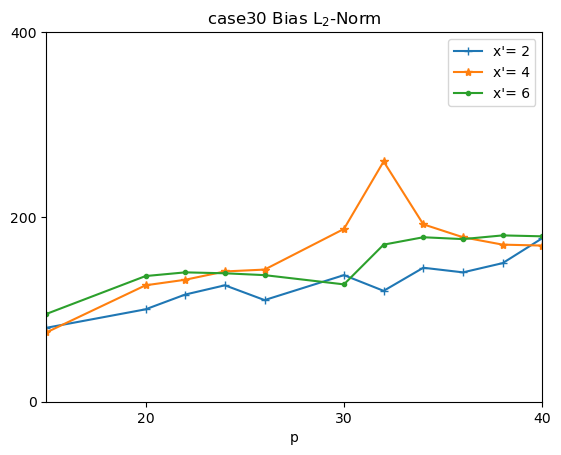}
    \label{fig:case30_bias}
    }
\subfloat{
    \includegraphics[width=.24\textwidth]{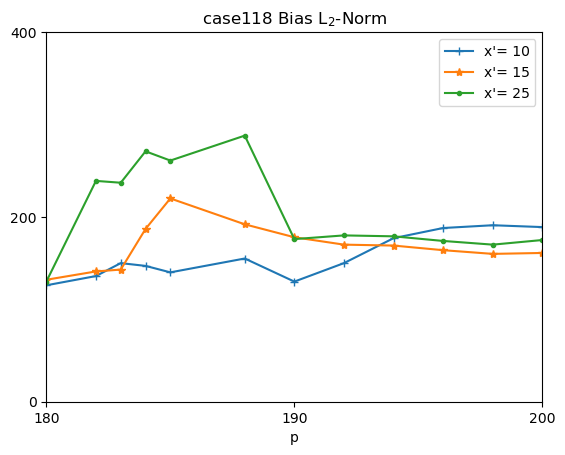}
    \label{fig:case118_bias}
    }
\subfloat{
    \includegraphics[width=.24\textwidth]{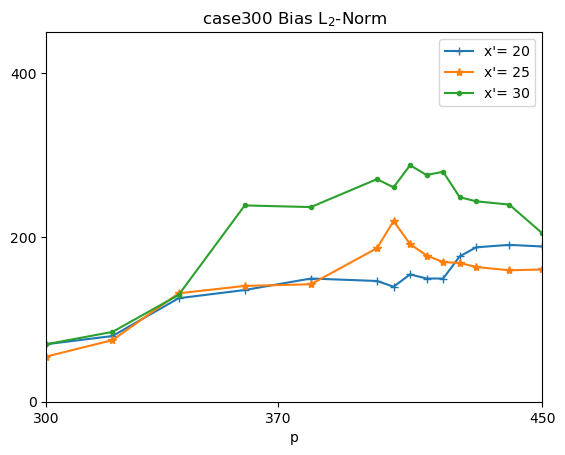}
    \label{fig:case300_bias}
    }
    \caption{Results for Bias L2 norm for bus case14, case30 and case118}
\end{figure*}

\begin{figure*}
\centering 
\subfloat{
    \includegraphics[width=.24\textwidth]{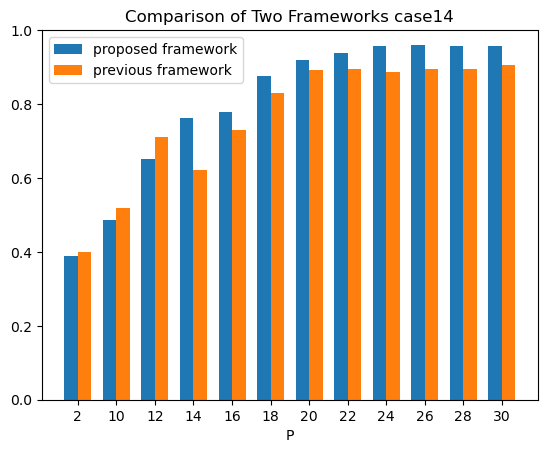}
    \label{fig:case14_com}
    }
\subfloat{
    \includegraphics[width=.24\textwidth]{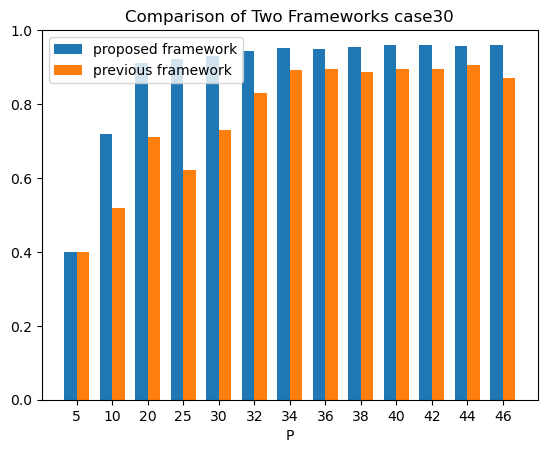}
    \label{fig:case30_com}
    }
\subfloat{
    \includegraphics[width=.24\textwidth]{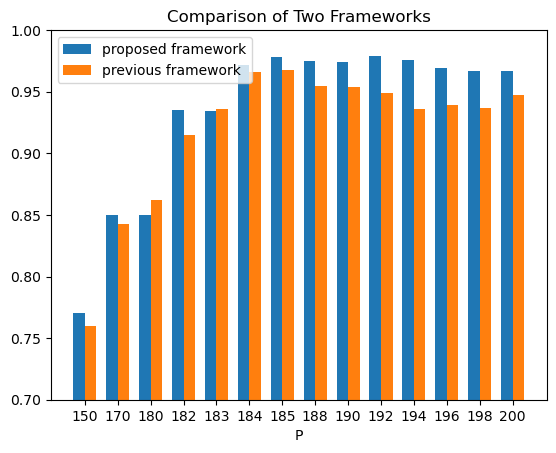}
    \label{fig:case118_com}
    }
\subfloat{
    \includegraphics[width=.24\textwidth]{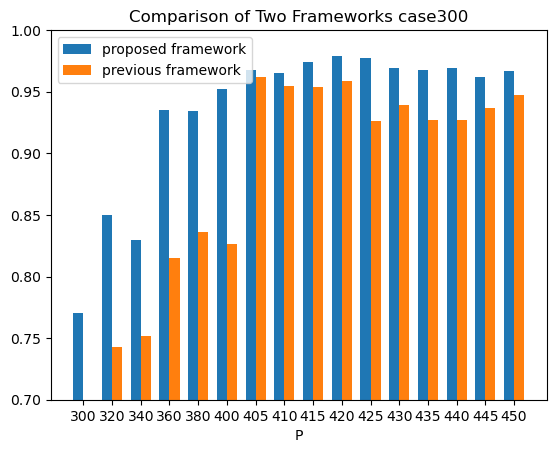}
    \label{fig:case300_com}
    }
    \caption{Detection accuracy comparison between two frameworks for bus case14, case30 and case118}
\end{figure*}

\begin{itemize}
    \item \textbf{Vanilla Attack}: The attackers are unaware of the presence of randomization layers, and the target model remains in its original network form.
    \item \textbf{Collective-Pattern Attack}: The attackers are aware of the presence of randomization layers. To better emulate the structures of defense models and achieve a more realistic representation, the target model is selected as the original network augmented with randomization layers, incorporating a diverse set of predefined patterns.
\end{itemize}

\subsection{Results}
In Figure 5, we observe the trend of the Bias L2-Norm, which plays a crucial role in bypassing FDIA detection. The Bias L2-Norm is a measure of the deviation or difference between the generated adversarial examples and the normal data distribution. To effectively bypass FDIA detection, attackers aim to achieve a relatively higher probability of successful evasion. This goal can be accomplished by obtaining a high-bias L2 norm. A higher Bias L2-Norm signifies a greater discrepancy between the adversarial examples and the expected normal behavior, making it more challenging for the detection algorithms to identify them as malicious. Comparing our random padding framework with the plain DNN, we find that our framework is capable of increasing the Bias L2-Norm. This increase in the Bias L2-Norm acts as a countermeasure against FDIA attacks that specifically target local marginal prices. By increasing the Bias L2-Norm, our random padding framework diminishes the effectiveness and performance of FDIA, thereby enhancing the system's resilience against such attacks. The three plots under Figure 5 demonstrate the changes in bias L2-norm under different attack scenarios for different bus test cases while increasing the padding size. In all attack scenarios for different bus test cases, the increasing padding size tends to have a positive effect on the bias L2-norm.

Figure 6 presents a comprehensive comparison between our proposed framework and a similar framework proposed in \cite{li2021towards} across various bus test case scenarios. The analysis specifically focuses on the impact of increasing padding sizes. Our proposed framework consistently delivers favorable results when compared to the framework in \cite{li2021towards}  with different padding sizes.
In the case of bus test case 14, our proposed framework demonstrates a notable increase in detection accuracy compared to the other framework. The accuracy gradually improves as the padding size increases, reaching its peak at a padding size of 24. Importantly, even with further increases in padding size, our proposed framework maintains consistently high accuracy. In contrast, the other framework exhibits detection accuracy fluctuations and, at certain points, experiences a decline in accuracy for the same bus test case 14.
For test case 30, our proposed framework exhibits higher detection accuracy compared to the other framework. Impressively, our proposed framework achieves the highest accuracy with a considerably lower number of padding sizes. Furthermore, as the padding size continues to increase, the detection accuracy of our proposed framework remains constant. This indicates the robustness and stability of our framework's performance across different padding sizes.
For bus test case 118, our proposed framework surpasses the other framework, demonstrating the most favorable results. In the other framework, the detection accuracy gradually deteriorates as the padding size increases up to a certain threshold. However, our proposed framework showcases a comparatively minimal decline pattern in detection accuracy for the same test case scenario. This highlights the superior performance and resilience of our proposed framework.
Overall, our proposed framework consistently achieves high levels of detection accuracy, showcasing its effectiveness in the context of intrusion detection. The comprehensive analysis presented in Figure 6 validates the superiority of our approach, offering a robust defense mechanism against potential threats and ensuring the overall security of the system.

\section{Conclusion}
Adversarial attacks like FIDA pose a significant threat even to the detection methods that include DNN models and ensuring security to physical level sensors is infeasible along with highly depends on specific hardware deployment. Therefore, in this paper, we study a defense mechanism that utilizes a randomization input reconstruction technique that helps to make the adversarial attacks less effective on the DNN-base FDIA detection frameworks. We evaluate our framework against nontrivial attacks that cause significant accuracy digression to the model's decision. Our easily deployable framework is compatible with any DNN model and shows favorable results on different IEEE bus test systems

\vspace{1cm}

\bibliographystyle{IEEEtran}
\bibliography{IEEEabrv,reference}

\end{document}